\documentclass[a4paper]{jpconf}
\usepackage{graphicx}
\usepackage[utf8]{inputenc}
\usepackage{amsmath}
\usepackage{verbatim}
\usepackage{color}
\usepackage{wrapfig}

\begin{document}
\title{Hints on the origin of the thermal hysteresis suppression in giant magnetocaloric thin films irradiatied with highly charged ions}

\author{S.~Cervera$^{1}$, M.~Trassinelli$^{1}$, M.~Marangolo$^{1}$, L.~Bernard Carlsson$^{1}$, M.~Eddrief$^{1}$, V.H.~Etgens$^{1,2,3}$, V.~Gafton$^{1,4}$, S.~Hidki$^{1}$, E.~Lamour$^{1}$, A.~Lévy$^{1}$, S.~Macé$^{1}$, C.~Prigent$^{1}$, J.-P.~Rozet$^{1}$, S.~Steydli$^{1}$, Y.~Zheng$^{1}$ and D.~Vernhet$^{1}$}
 
\address{$^1$ Sorbonne Universités, UPMC Univ Paris 06, CNRS-UMR 7588, Institut des NanoSciences de Paris, F-75005, Paris, France} 
\address{$^2$ Université Versailles St-Quentin, LISV, Bâtiment Boucher, Pôle scientifique et technologique de Vélizy, 10-12 avenue de l’Europe, 78140 Vélizy, France}
\address{$^3$ Institut VEDECOM, 77 rue des Chantiers, 78000 Versailles, France}
\address{$^4$ Alexandru Ioan Cuza University, Faculty of Physics, 11 Carol I Blv, Iasi 700506, Romania}

\ead{sophie.cervera@insp.jussieu.fr, martino.trassinelli@insp.jussieu.fr}

\begin{abstract}
In a recent experiment we demonstrated the possibility to suppress the thermal hysteresis of the phase transition in giant magnetocaloric MnAs thin film by interaction with slow highly charged ions (Ne$^{9+}$ at 90~keV) \cite{Trassinelli2014}. This phenomenon has a major impact for possible applications in magnetic refrigeration and thus its reproducibility and robustness are of prime importance.
Here we present some new investigations about the origin and the nature of the irradiation-induced defects responsible for the thermal hysteresis suppression. Considering in particular two samples that receive different ion fluences (two order of magnitude of difference), we investigate the reliability of this process. The stability of the irradiation-induced defects with respect to a soft annealing is studied by X-ray diffraction and magnetometry measurements, which provide some new insights on the mechanisms involved.
\end{abstract}

\section{Introduction}
Conventional magnetic materials increase their temperature when they are placed in a magnetic field and cool down when they are removed. This phenomenon is known as the magnetocaloric effect (MCE). 
The discovery of compounds with a giant magnetocaloric effect (GMCE) close to room temperature pushed the development of magnetic refrigeration as an environmentally friendly and energy-efficient technology alternative to gas compression refrigeration commonly employed for everyday applications. Up to now, the practical application of GMCE materials is blocked by the fact that these compounds are also characterised by a magnetic transition of first-order \cite{Li2012,Roy2013}. This type of transition suffers intrinsically from large thermal hysteresis making their use in thermal machine inefficient.
In the last decades many but mostly unsuccessful efforts have been done for eliminating the thermal hysteresis keeping the GMCE properties. 
These investigations explored various GMCE materials of different chemical composition and doping with additional elements \cite{Pecharsky1997a,Pecharsky1997,Wada2001,Provenzano2004,Krenke2005,Bruck2005,Wada2005,deCampos2006,Rocco2007,Sun2008,Bruck2008,Cui2009,Trung2009,Xu2010,Gutfleisch2011,Bratko2012,Dung2012,Teixeira2012,Franco2012,Sandeman2012,Smith2012,Caron2013,Caron2013a,Moya2014}, but also the application of external strain to bulk materials \cite{Liu2012} or on thin films  \cite{Mosca2008,Duquesne2012,Moya2013,Marangolo2014,Teichert2015}.

Quite recently, our group has developed a new approach to suppress the thermal hysteresis in GMCE materials based on the interaction with highly charged ions \cite{Trassinelli2014,Trassinelli2013}.
We have demonstrated that the irradiation of manganese arsenide (MnAs) thin films (150~nm thick)  with a moderate fluence (about 10$^{13}$ ions cm$^{-2}$) of slow highly charged ions (Ne$^{9+}$ with a kinetic energy of 90~keV) can completely suppress the thermal hysteresis keeping the other magnetic and structural properties almost unchanged.
In particular, the large refrigerant power of samples is preserved.

The possibility to suppress the thermal hysteresis seems to be connected to the nature of the transition related to the GMCE.
The MnAs is characterised by a first-order magneto-structural phase transition from a $\alpha$-phase hexagonal and ferromagnetic to a $\beta$-phase  orthorhombic and non-ferromagnetic (paramagnetic or antiferromagnetic).
Compared to bulk materials, in MnAs thin films, the strain of the substrate disturbs the phase transition leading to the $\alpha - \beta$ phase coexistence but keeping the GMCE.

\begin{wrapfigure}{RT}{0.35\textwidth} 
\centering
\includegraphics[width=0.3\textwidth]{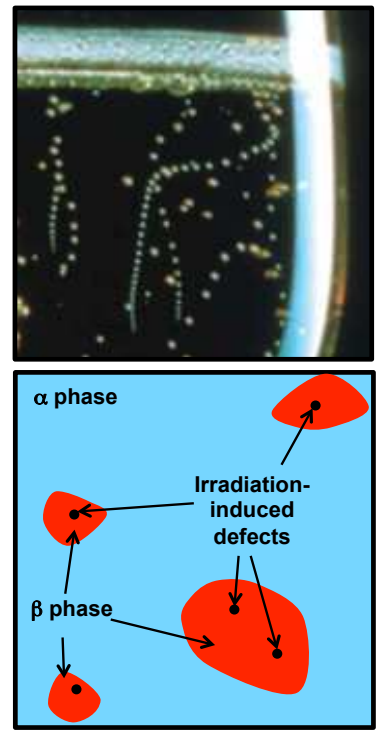} \\
\caption{\label{fig:nucleation} Up: Chains of CO$_2$ vapour bubbles in proximity of glass defects \cite{Liger-Belair2008}. 
Down: Analog artistic view of the nucleation of MnAs $\beta$ phase in $\alpha$ phase in proximity of the irradiation-induced defects.}
\end{wrapfigure}

As in all first-order transitions, and especially in structural phase transitions like liquid--solid or liquid--vapour, a nucleation energy barrier has to be overcome for the crossing of one phase into the other phase.
This phenomenon is responsible for phase metastability, with possibility of supercooled or superheated states as well as thermal hysteresis. 
As an example, in champagne or beer, at room temperature (or better fresh) and pressure, while the equilibrium state correspond to the vapour state, we can have the presence of supersaturated dissolved CO$_2$ molecules in the liquid.
In a glass of champagne or beer, the recipient surface defects or small grains of dust facilitate the creation of CO$_2$ vapour bubbles from the supersaturated molecules dissolved in the liquid \cite{Liger-Belair2008,Shafer1991} with the creation of typical chains of small bubbles arising from a precise point in the glass (Fig.~\ref{fig:nucleation} up).
The local small radii of surface defects or dust grains lower the structure energy barrier facilitating the nucleation.

Similarly, the defects induced by the ion impact in MnAs act as seeds during the transition for the nucleation of one phase into the other (Fig.~\ref{fig:nucleation} down).
This defects leads the permanent suppression of the thermal hysteresis (stable in time) but their exact nature is not identified.

In the considered velocity regime, the ion--solid interaction is dominated by the nuclear stopping power with binary collisions between projectile and target atoms. 
From these collisions point-like defects are created: implantation of projectile (Ne) atoms or displacement of target (Mn or As) atoms between the crystal planes and production of atomic vacancies in the target crystal lattice. 
Consequently, one can expect that some of these defects may recover with thermal treatment which can cause recombination of displaced  atoms and vacancies or even the expulsion of deposited projectile atoms.
In this article we investigate the effect of annealing of irradiated samples by studying of the magnetic and structural properties of irradiated samples before and after thermal treatment.
This characterisation allows to test the stability and robutness of this process with temperature, but above all it provides some hints for the determination of the nature of the irradiation-induced defects responsible for the thermal hysteresis elimination.

The article is organised as follows.
In next section, we describe the experimental methods of production, irradiation and characterisation of the samples. 
In section 3 the results are discussed and a conclusion is presented section 4.

\section{Experimental methods}
All the monocrystal MnAs epilayers (with a thickness of $150 \pm 10$~nm) investigated here are issued from the same growth obtained by molecular beam epitaxy (MBE) on GaAs(001) substrate. 
Details on the growth process can be found in Ref.~\cite{Breitwieser2009}.

The different samples are irradiated with a Ne$^{9+}$  ion beam at the SIMPA facility \cite{Gumberidze2010} (French acronym for highly charged ion source of Paris). 
The ion kinetic energy is set to 90~keV (4.5~keV/u) with an incidence angle of 60$^\circ$.
In this condition, the average penetration depth of the ions corresponds to the half-thickness of the MnAs film\cite{Ziegler2008}, without penetration into the substrate,  with a consequent maximisation of the irradiation effect \cite{Zhang2004}.
In the present paper we consider two irradiated samples that received a fluence of $1.5 \times 10^{13}$ and  $1.6 \times 10^{15}$ ions cm$^{-2}$, respectively.
In the following these samples are referred with the adjective either \textit{low-} or \textit{high-fluence}.
More details about the irradiation process can be found in Refs.~\cite{Trassinelli2014,Trassinelli2013}.

As discussed before, for studying the irradiation-induced defects, the irradiated samples are annealed under helium atmosphere at a temperature of 400~K for about two hours. 
At this temperature no modifications are induced in non-irradiated MnAs thin films \cite{Daweritz2006}.
Before and after annealing (but also before the irradiation), the samples properties are characterised by X-ray diffraction (XRD) and magnetometry.

The XRD data are obtained at room temperature, where the two phases of MnAs are normally visible. 
We use two different instruments, both using the CuK$\alpha$ emission: a PANalytical XPert MRD diffractometer (without monochromator) and a Rigaku Smartlab diffractometer with a copper rotating anode and a two-reflection monochromator.

Measurements of the magnetisation in different conditions (temperature and external magnetic field) are obtained by a vibrating sample magnetometer 
(VSM, in the Quantum Design PPMS 9T) and a SQUID magnetometer (Quantum Design MPMS-XL).
For the temperature dependency of the sample magnetisation, we apply an external field of $H=1$~T and vary the temperature between 150 and 350~K with a rate of $\pm 2$~K/min.
\textit{N.B.} up to 350~K, the irradiated samples remain stable and no changes on their characteristics are noticed.
When comparing the results of the two magnetometers, we realised that the VSM was overestimating and underestimating the sample temperature during the heating and the cooling, respectively, due to a temperature sensor placed far from the sample. 
No similar problem is present in the SQUID magnetometer.

Since some data have been obtained exclusively with the VSM before the annealing, we correct the corresponding VSM measurements from the bias of the temperature reading. 
The correction is obtained by a comparison of VSM and SQUID magnetic moment measurements obtained in identical conditions (same samples, temperature change rate and applied field).

Information on the magnetic cycle, and more precisely on the coercivity field (the half width of the magnetic cycle) at different temperatures are extracted from the hysteresis curves of the magnetisation $M$ with a variable magnetic field $H$ between -1 and 1 T. 
Before each cycle, the samples are depolarised at $T=350$~K and $H=0$, and then brought to the selected temperature. 
This allows for cancelling their magnetic and thermal history to avoid artefacts in the measurements, as discussed in Refs. \cite{Basso2007,Caron2009,Tocado2009}.

\begin{figure}[t]
\centering
\includegraphics[width=0.49\textwidth]{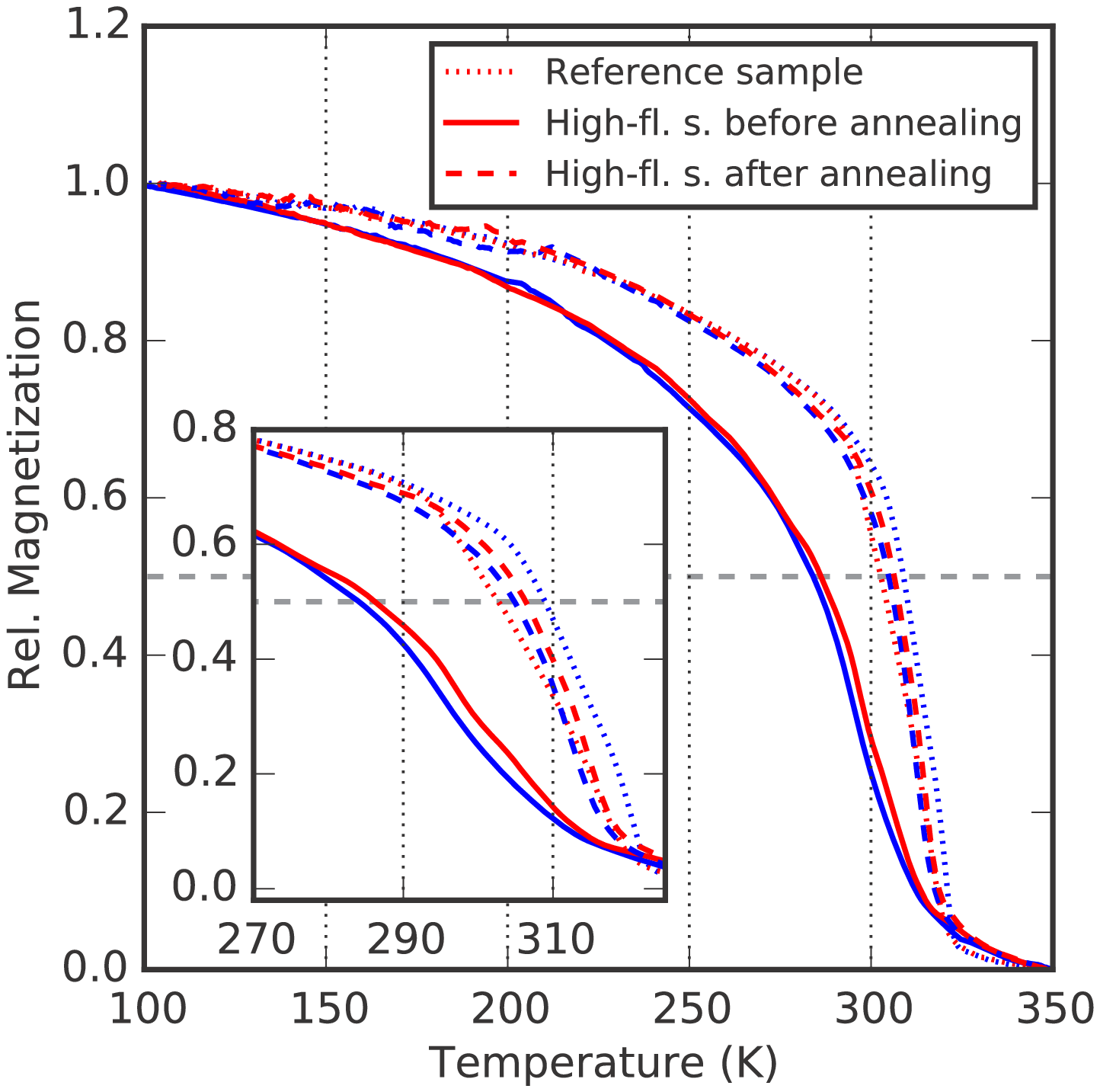}
\includegraphics[width=0.49\textwidth]{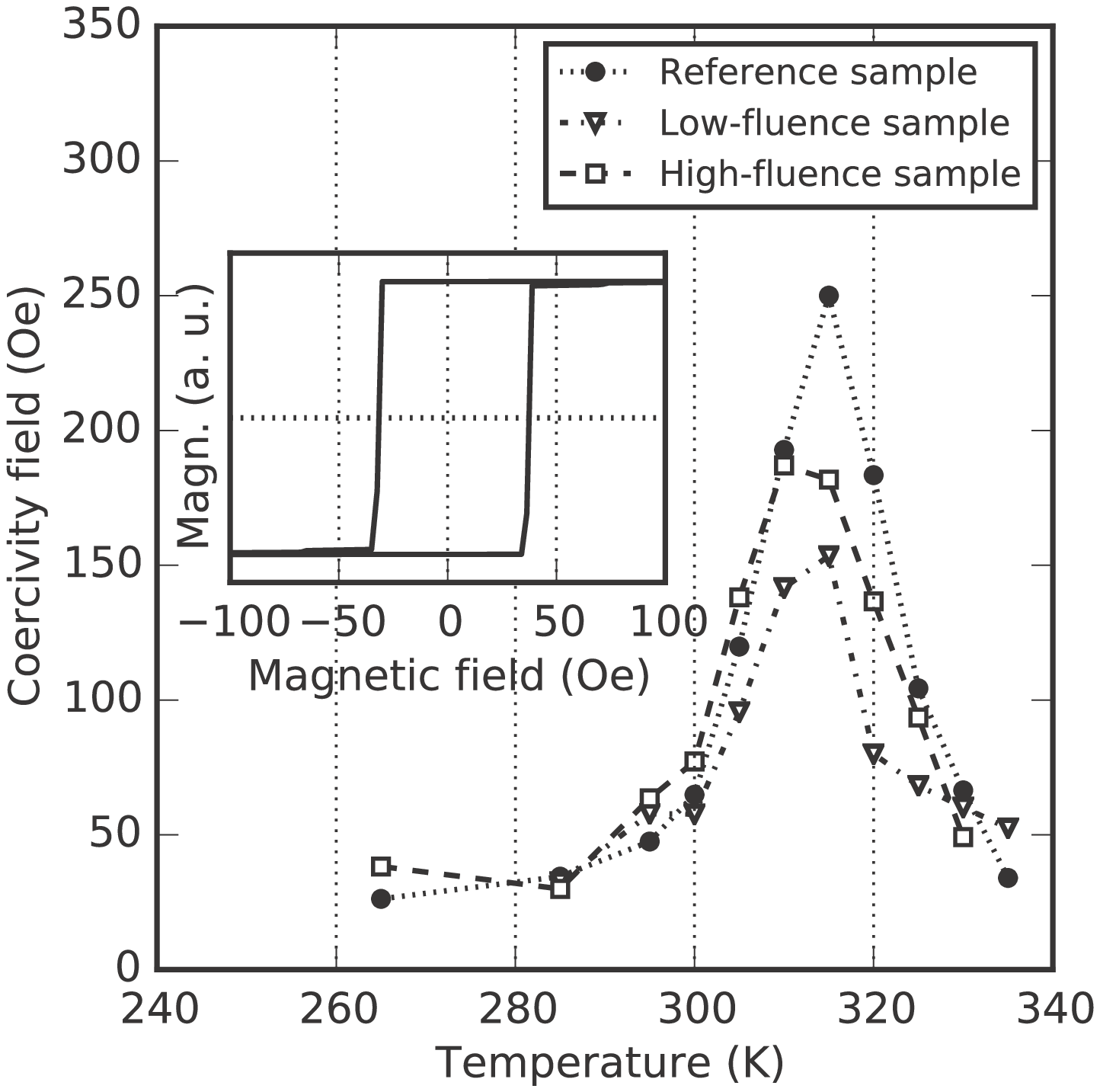}
\caption{\label{fig:repairing} (colour online) Left: Magnetisation as a function of temperature for the reference and for the \textit{high-fluence} samples before and after annealing . Data obtained by a temperature increase (from colder temperatures) and decrease (from hotter temperatures) are presented in blue and red, respectively. The horizontal dashed lines indicates half of the saturation magnetisation value from which the transition temperature $T_c$ is calculated.
Right: Coercivity field at different temperatures of the reference and the two irradiated samples before annealing.
In the inset we present a typical magnetic cycle (relative to $T=285$~K) from which the coercivity values are extracted.}
\end{figure}

\section{Results and discussion}
As reported in Ref.~\cite{Trassinelli2014}, the \textit{low-fluence} sample does not present large magnetic properties differences with the non-irradiated reference except on the suppression of the thermal hysteresis. 
In particular, the temperature dependency of the magnetisation $M(T)$ results to be the same.
As it can be observed in Fig.~\ref{fig:repairing} left, this is not the case for the \textit{high-fluence} sample.  
Here the transition temperature (or critical temperature, defined as the temperature $T_c$ where we have a magnetisation half of the saturation magnetisation, $M(T_c)=M_{sat}/2$) is shifted to $285~$K from the original value $T_c = 307$~K.
A similar behaviour is observed in MnAs doped with Sb \cite{Wada2001} or Fe \cite{deCampos2006}.
More generally, the slope of the  $M(T)$ curve is significantly different in the proximity of $T_c$. 
This corresponds to a magnetisation decreasing at a fixed temperature, similarly to what has been observed at room temperature in ferromagnetic thin films (without first-order transitions) irradiated with singly charged ions \cite{Chappert1998,Kaminsky2001,Zhang2003,Fassbender2006,Fujita2010,Cook2011,Oshima2013}. 
In these experiments the magnetisation reduction is interpreted by the amorphisation or structural phase changes of the films produced by ion collisions.
As in the \textit{low-fluence} sample, for the \textit{high-fluence} sample we get suppression of the thermal hysteresis as well, with some indications of its possible inversion.

\begin{wrapfigure}{RT}{0.4\textwidth} 
\includegraphics[width=0.4\textwidth]{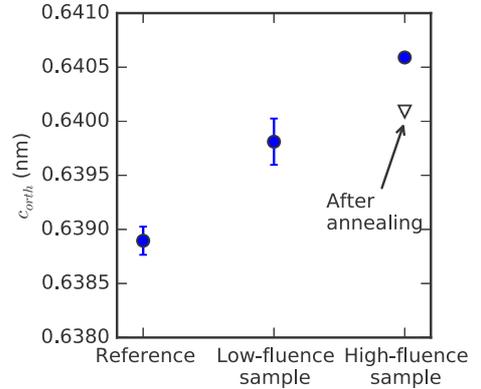} 
\caption{\label{fig:XRD} $c_\text{orth}$ values for the different samples extracted from XRD measurements.}
\end{wrapfigure}  

For both irradiated samples, the coercivity values at different temperatures exhibit similar results compared to the reference measurements (see Fig.~\ref{fig:repairing} right).
This indicates that, differently to other past experiments with singly charged ion discussed above, the ferromagnetic $\alpha$ phase of MnAs, responsible for the coercivity at low temperature,  is not perturbed by highly charged ion collisions and/or implantation.
Neither is perturbed the interplay between $\alpha$ and $\beta$ phase coexistence during the transition, which is responsible for the characteristic peak of the coercivity during the transition \cite{Steren2006}.

As already discussed in \cite{Trassinelli2014}, a visible effect of the ion irradiations is present in the XRD data. 
The ion bombardment introduces, in fact, additional constraints between MnAs $\alpha$ and $\beta$ phases. 
In practice the $\alpha$ and $\beta$ lattice spacing get closer with a complete merging of the XRD peaks for the \textit{high-fluence} sample where the two phases are not distinguishable anymore.
However, the characteristic full-width-at-half-maximum (FWHM) of the diffraction peaks does not change significantly as a function of the received ion fluence. 

Similarly to the coercivity field measurement, this indicates that a possible amorphisation of the crystalline phases induced by the irradiation has to be excluded.
We can conclude therefore that, in our conditions, the irradiation-induced defects do not perturb too much the sample structure.

From the XRD data, the orthorhombic lattice constant $c_\text{orth}$ value can be analysed as a function of the ion fluence. 
We consider here the arithmetical average of $c_\text{orth}$ of the two phases, which changes slightly during the transition even if the fraction between the two phase changes radically \cite{Adriano2006}.
As visible in Fig.~\ref{fig:XRD}, higher irradiation fluence values correspond to higher $c_\text{orth}$.
More precisely the \textit{low-} and the \textit{high-fluence} samples are found to have a value of $c_\text{orth}$ 0.14\% and 0.27\% larger than the reference one.

After the annealing, no noticeable changes have been detected in the \textit{low-fluence} sample. In opposite, the \textit{high-fluence} sample is subject to a series of modifications. As it can be observed in Fig.~\ref{fig:repairing} left, the most evident change is the dependency of the magnetisation with the temperature that, after the annealing, becomes very similar to the reference sample.  
The thermal hysteresis is kept suppressed, indicating that the defects responsible for it are quite robust and cannot easily be recovered with the annealing. 

In the correspondence, the relative $c_\text{orth}$ value changes  from 0.27\% before annealing to 0.19\% after. 
This contraction can be the result of the evacuation of the interstitial Ne atoms but also the lattice relaxation of the vacancy defects.

Data from XRD and the magnetometers demonstrate together that some irradiation-induced defects are certainly repaired. 
During the annealing, the temperature raise leads to an increase of the atom mobility in the sample up to a possible ejection of the projectile atoms \cite{Guo2015}.
A replacement between target and projectile atoms are unlikely due to the noble nature of Neon that is loath to create chemical bonds.
In opposite, some of the pairs of interstitial target atom and vacancy created during the collision with the projectile (Frenkel pairs) can be recombined. 
This is the most likely mechanism responsible for the observed changes on $M(T)$ and $c_\text{orth}$.
The exact nature of the defects that induce the thermal hysteresis suppression is still ambiguous, but could be caused by pairs of interstitial atoms and vacancies too far to be recombined or by the presence of implanted projectile atoms that are not easily removed.

\section{Conclusion}
We present some new investigations about the origin and the nature of the  irradiation-induced defects responsible for the thermal hysteresis suppression in MnAs thin films bombarded with highly charged ions (Ne$^{9+}$).
Considering in particular two samples that receive different ion fluences (two order of magnitude of difference), we investigate the stability of these new defects and their robustness with respect to a soft annealing process.
Comparing X-ray diffraction and magnetometry data obtained before and after the annealing, we demonstrate that a partial recovery of the irradiation-induced defects is possible for samples that received \textit{high-fluence} of ions.
In both samples the thermal hysteresis is maintained suppressed demonstrating that its elimination via the highly charged ion irradiation is stable and robust with respect to external factors making this technique suitable for modifying giant magnetocaloric materials for magnetic refrigeration applications.
The exact nature of the defects induced by the ion irradiation and responsible for the thermal hysteresis suppression is still not completly elucidated and additional investigation are required.

\ack
This work was partially supported by French state funds managed by the ANR within the Investissements d'Avenir programme under reference ANR-11-IDEX-0004-02, and more specifically within the framework of the Cluster of Excellence MATISSE led by Sorbonne Universités. 
V. Gafton acknowledges support from the strategic grant  POSDRU/159/1.5/S/137750, Project, ``Doctoral and Post-doctoral programs support for increased competitiveness in Exact Sciences research'' co financed by the European Social Found within the Sectoral Operational Program Human Resources Development 2007-2013

\section*{References}

\bibliography{Cervera_ICPEAC2015}

\providecommand{\newblock}{}
\begin{thebibliography}{10}
\expandafter\ifx\csname url\endcsname\relax
  \def\url#1{{\tt #1}}\fi
\expandafter\ifx\csname urlprefix\endcsname\relax\def\urlprefix{URL }\fi
\providecommand{\eprint}[2][]{\url{#2}}

\bibitem{Trassinelli2014}
Trassinelli M, Marangolo M, Eddrief M, Etgens V~H, Gafton V, Hidki S, Lacaze E,
  Lamour E, Prigent C, Rozet J~P, Steydli S, Zheng Y and Vernhet D 2014 {\em
  Appl. Phys. Lett.\/} {\bf 104} 081906

\bibitem{Li2012}
Li B, Meng H, Ren W and Zhang Z 2012 {\em Eur. Phys. Lett.\/} {\bf 97} 57002

\bibitem{Roy2013}
Roy S~B 2013 {\em J. Phys. Condens. Matter\/} {\bf 25} 183201

\bibitem{Pecharsky1997a}
Pecharsky V~K and Gschneidner J~K~A 1997 {\em Phys. Rev. Lett.\/} {\bf 78}
  4494--4497

\bibitem{Pecharsky1997}
Pecharsky V~K and Gschneidner J~K~A 1997 {\em Appl. Phys. Lett.\/} {\bf 70}
  3299--3301

\bibitem{Wada2001}
Wada H and Tanabe Y 2001 {\em Appl. Phys. Lett.\/} {\bf 79} 3302--3304

\bibitem{Provenzano2004}
Provenzano V, Shapiro A~J and Shull R~D 2004 {\em Nature\/} {\bf 429} 853--857

\bibitem{Krenke2005}
Krenke T, Duman E, Acet M, Wassermann E~F, Moya X, Manosa L and Planes A 2005
  {\em Nature Mater.\/} {\bf 4} 450--454

\bibitem{Bruck2005}
Brück E 2005 {\em J. Phys. D\/} {\bf 38} R381

\bibitem{Wada2005}
Wada H and Asano T 2005 {\em J. Magn. Magn. Mater.\/} {\bf 290} 703--705

\bibitem{deCampos2006}
de~Campos A, Rocco D~L, Carvalho A~M~G, Caron L, Coelho A~A, Gama S, da~Silva
  L~M, Gandra F~C~G, dos Santos A~O, Cardoso L~P, von Ranke P~J and de~Oliveira
  N~A 2006 {\em Nature Mater.\/} {\bf 5} 802--804

\bibitem{Rocco2007}
Rocco D~L, de~Campos A, Carvalho A~M~G, Caron L, Coelho A~A, Gama S, Gandra
  F~C~G, dos Santos A~O, Cardoso L~P, von Ranke P~J and de~Oliveira N~A 2007
  {\em Appl. Phys. Lett.\/} {\bf 90} 242507--3

\bibitem{Sun2008}
Sun N~K, Cui W~B, Li D, Geng D~Y, Yang F and Zhang Z~D 2008 {\em Appl. Phys.
  Lett.\/} {\bf 92} 072504--3

\bibitem{Bruck2008}
Brück E, Tegus O, Cam~Thanh D~T, Trung N~T and Buschow K~H~J 2008 {\em Int. J.
  Refrig.\/} {\bf 31} 763--770

\bibitem{Cui2009}
Cui W~B, Liu W, Liu X~H, Guo S, Han Z, Zhao X~G and Zhang Z~D 2009 {\em J.
  Alloys Compd.\/} {\bf 479} 189--192

\bibitem{Trung2009}
Trung N~T, Ou Z~Q, Gortenmulder T~J, Tegus O, Buschow K~H~J and Bruck E 2009
  {\em Appl. Phys. Lett.\/} {\bf 94} 102513--3

\bibitem{Xu2010}
Xu P~F, Nie S~H, Meng K~K, Wang S~L, Chen L and Zhao J~H 2010 {\em Appl. Phys.
  Lett.\/} {\bf 97} --

\bibitem{Gutfleisch2011}
Gutfleisch O, Willard M~A, Brück E, Chen C~H, Sankar S~G and Liu J~P 2011 {\em
  Adv. Mater.\/} {\bf 23} 821--842

\bibitem{Bratko2012}
Bratko M, Morrison K, de~Campos A, Gama S, Cohen L~F and Sandeman K~G 2012 {\em
  Appl. Phys. Lett.\/} {\bf 100} --

\bibitem{Dung2012}
Dung D~D, Tuan D~A, Van~Thiet D, Shin Y and Cho S 2012 {\em J. Appl. Phys.\/}
  {\bf 111} 07C310--3

\bibitem{Teixeira2012}
Teixeira C~S, Krautz M, Moore J~D, Skokov K, Liu J, Wendhausen P~A~P and
  Gutfleisch O 2012 {\em J. Appl. Phys.\/} {\bf 111} 07A927

\bibitem{Franco2012}
Franco V, Blázquez J, Ingale B and Conde A 2012 {\em Annu. Rev. Mater. Res.\/}
  {\bf 42} 305--342

\bibitem{Sandeman2012}
Sandeman K~G 2012 {\em Scripta Mat.\/} {\bf 67} 566--571

\bibitem{Smith2012}
Smith A, Bahl C~R~H, Bjørk R, Engelbrecht K, Nielsen K~K and Pryds N 2012 {\em
  Adv. Energy Mater.\/} {\bf 2} 1288--1318

\bibitem{Caron2013}
Caron L, Hudl M, Höglin V, Dung N~H, Gomez C~P, Sahlberg M, Brück E,
  Andersson Y and Nordblad P 2013 {\em Phys. Rev. B\/} {\bf 88} 094440

\bibitem{Caron2013a}
Caron L, Miao X~F, Klaasse J~C~P, Gama S and Brück E 2013 {\em Appl. Phys.
  Lett.\/} {\bf 103} 112404

\bibitem{Moya2014}
Moya X, Kar-Narayan S and Mathur N~D 2014 {\em Nature Mater.\/} {\bf 13}
  439--450

\bibitem{Liu2012}
Liu J, Gottschall T, Skokov K~P, Moore J~D and Gutfleisch O 2012 {\em Nature
  Mater.\/} {\bf 11} 620--626

\bibitem{Mosca2008}
Mosca D~H, Vidal F and Etgens V~H 2008 {\em Phys. Rev. Lett.\/} {\bf 101}
  125503

\bibitem{Duquesne2012}
Duquesne J~Y, Prieur J~Y, Canalejo J~A, Etgens V~H, Eddrief M, Ferreira A~L and
  Marangolo M 2012 {\em Phys. Rev. B\/} {\bf 86} 035207

\bibitem{Moya2013}
Moya X, Hueso L~E, Maccherozzi F, Tovstolytkin A~I, Podyalovskii D~I, Ducati C,
  Phillips L~C, Ghidini M, Hovorka O, Berger A, Vickers M~E, Defay E, Dhesi S~S
  and Mathur N~D 2013 {\em Nature Mater.\/} {\bf 12} 52--58

\bibitem{Marangolo2014}
Marangolo M, Karboul-Trojet W, Prieur J~Y, Etgens V~H, Eddrief M, Becerra L and
  Duquesne J~Y 2014 {\em Appl. Phys. Lett.\/} {\bf 105} 162403

\bibitem{Teichert2015}
Teichert N, Kucza D, Yildirim O, Yuzuak E, Dincer I, Behler A, Weise B, Helmich
  L, Boehnke A, Klimova S, Waske A, Elerman Y and Hütten A 2015 {\em Phys.
  Rev. B\/} {\bf 91} 184405

\bibitem{Trassinelli2013}
Trassinelli M, Gafton V~E, Eddrief M, Etgens V~H, Hidki S, Lacaze E, Lamour E,
  Luo X, Marangolo M, Mérot J, Prigent C, Reuschl R, Rozet J~P, Steydli S and
  Vernhet D 2013 {\em Nucl. Instrum. Methods B\/} {\bf 317} 154--158

\bibitem{Liger-Belair2008}
Liger-Belair G, Polidori G and Jeandet P 2008 {\em Chem. Soc. Rev.\/} {\bf 37}
  2490--2511

\bibitem{Shafer1991}
Shafer N~E and Zare R~N 1991 {\em Phys. Today\/} {\bf 44} 48--52

\bibitem{Breitwieser2009}
Breitwieser R, Vidal F, Graff I~L, Marangolo M, Eddrief M, Boulliard J~C and
  Etgens V~H 2009 {\em Phys. Rev. B\/} {\bf 80} 045403

\bibitem{Gumberidze2010}
Gumberidze A, Trassinelli M, Adrouche N, Szabo C~I, Indelicato P, Haranger F,
  Isac J~M, Lamour E, Le~Bigot E~O, Merot J, Prigent C, Rozet J~P and Vernhet D
  2010 {\em Rev. Sci. Instrum.\/} {\bf 81} 033303--10

\bibitem{Ziegler2008}
Ziegler J~F, Biersack J~P and Ziegler M~D 2008 {\em Stopping and Range of Ions
  in Matter\/} (SRIM Company)

\bibitem{Zhang2004}
Zhang K, Lieb K~P, Müller G~A, Schaaf P, Uhrmacher M and Münzenberg M 2004
  {\em Eur. Phys. J. B\/} {\bf 42} 193--204

\bibitem{Daweritz2006}
Däweritz L 2006 {\em Rep. Prog. Phys.\/} {\bf 69} 2581

\bibitem{Basso2007}
Basso V, Sasso C~P and LoBue M 2007 {\em J. Magn. Magn. Mater.\/} {\bf 316}
  262--268

\bibitem{Caron2009}
Caron L, Ou Z~Q, Nguyen T~T, Cam~Thanh D~T, Tegus O and Brück E 2009 {\em J.
  Magn. Magn. Mater.\/} {\bf 321} 3559--3566

\bibitem{Tocado2009}
Tocado L, Palacios E and Burriel R 2009 {\em J. Appl. Phys.\/} {\bf 105} --

\bibitem{Chappert1998}
Chappert C, Bernas H, Ferré J, Kottler V, Jamet J~P, Chen Y, Cambril E,
  Devolder T, Rousseaux F, Mathet V and Launois H 1998 {\em Science\/} {\bf
  280} 1919--1922

\bibitem{Kaminsky2001}
Kaminsky W~M, Jones G~A~C, Patel N~K, Booij W~E, Blamire M~G, Gardiner S~M, Xu
  Y~B and Bland J~A~C 2001 {\em Appl. Phys. Lett.\/} {\bf 78} 1589--1591

\bibitem{Zhang2003}
Zhang K, Gupta R, Lieb K~P, Luo Y, Müller G~A, Schaaf P and Uhrmacher M 2003
  {\em Eur. Phys. Lett.\/} {\bf 64} 668

\bibitem{Fassbender2006}
Fassbender J, von Borany J, Mücklich A, Potzger K, Möller W, McCord J,
  Schultz L and Mattheis R 2006 {\em Phys. Rev. B\/} {\bf 73} 184410

\bibitem{Fujita2010}
Fujita N, Kosugi S, Saitoh Y, Kaneta Y, Kume K, Batchuluun T, Ishikawa N,
  Matsui T and Iwase A 2010 {\em J. Appl. Phys.\/} {\bf 107} 09E302

\bibitem{Cook2011}
Cook P~J, Shen T~H, Grundy P~J, Im M~Y, Fischer P, Morton S~A and Kilcoyne
  A~L~D 2011 {\em J. Appl. Phys.\/} {\bf 109} 063917--5

\bibitem{Oshima2013}
Oshima D, Kato T, Iwata S and Tsunashima S 2013 {\em IEEE Trans. Magn.\/} {\bf
  49} 3608--3611

\bibitem{Steren2006}
Steren L~B, Milano J, Garcia V, Marangolo M, Eddrief M and Etgens V~H 2006 {\em
  Phys. Rev. B\/} {\bf 74} 144402

\bibitem{Adriano2006}
Adriano C, Giles C, Couto O~D~D, Brasil M~J~S~P, Iikawa F and Daweritz L 2006
  {\em Appl. Phys. Lett.\/} {\bf 88} 151906--3

\bibitem{Guo2015}
Guo H, Dong S, Rack P, Budai J, Beekman C, Gai Z, Siemons W, Gonzalez C~M,
  Timilsina R, Wong A~T, Herklotz A, Snijders P~C, Dagotto E and Ward T~Z 2015
  {\em Phys. Rev. Lett.\/} {\bf 114} 256801

\end{thebibliography}
\bibliographystyle{iopart-num}

\end{document}